\documentclass[a4paper]{mem}
\usepackage{natbib}
\usepackage{graphicx}
\usepackage[a4paper]{hyperref}
\idline{1}{1}
\newcommand{\hst}{{\sl HST}}
\newcommand{\wcen}{{$\omega$~Cen}}
\begin{document}
\title{The double main sequence of Omega Centauri
\thanks{Based on observations with the NASA/ESA
 {\it Hubble Space Telescope}, obtained at the Space Telescope Science
 Institute, which is  operated by AURA,  Inc., under NASA contract NAS
 5-26555.}
}
 \author{L.\ R.\ Bedin \inst{1}, G.\ Piotto \inst{1}, J.\ Anderson \inst{2},\\
          I.\ R.\ King \inst{3}, S.\ Cassisi \inst{4}, 
          \and
          Y.\ Momany \inst{1}\fnmsep
%
%
}
%
%
\offprints{L.\ R.\ Bedin}
\mail{vic.\ Osservatorio 2, 35122 Padova }
\institute{Dip. di Astronomia, Univ. di Padova,
  vic.\ Osservatorio 2, 35122 Padova, Italia,
\email{bedin-piotto-momany@pd.astro.it}\\ 
\and  Dept.\ of Physics \& Astronomy, MS 108, Rice Univ., 
  6100 Main Street, Houston, TX 77005, USA,
\email{jay@eeyore.rice.edu}\\ 
\and Astronomy Dept., Univ.\ of Washington, Box 351580, 
  Seattle, WA 98195-1580, USA, 
\email{king@astro.washington.edu}\\
\and Osservatorio Astronomico di Collurania, via M. Maggini,
64100 Teramo, Italia,
\email{cassisi@astrte.te.astro.it}
}
   \abstract{Recent, high precision  photometry of Omega Centauri, the
   biggest  Galactic globular  cluster,  has been  obtained with  {\em
   Hubble  Space  Telescope}  (\hst).   The  color  magnitude  diagram
   reveals an  unexpected bifurcation of  colors in the  main sequence
   (MS).   The  newly  found  double  MS, the  multiple  turnoffs  and
   subgiant branches, and other sequences discovered in the past along
   the red  giant branch of this  cluster add up to  a fascinating but
   frustrating puzzle.   Among the possible explanations  for the blue
   main sequence  an anomalous  overabundance of helium  is suggested.
   The hypothesis will be tested with a set of FLAMES@VLT data we have
   recently obtained  (ESO DDT program), and  with forthcoming ACS@HST
   images.  
 }
   \authorrunning{L.\ R.\ Bedin et al.}
   \titlerunning{The Double MS of Omega Centauri}
   \maketitle
%

\section{Introduction}

Several properties  of ~Omega  ~Centauri (\wcen), including  the large
spread in  metallicity and the large  mass, make it  a peculiar object
among Galactic globular clusters (GGCs).

   \begin{figure*}
   \centering
   \resizebox{\hsize}{!}{\includegraphics{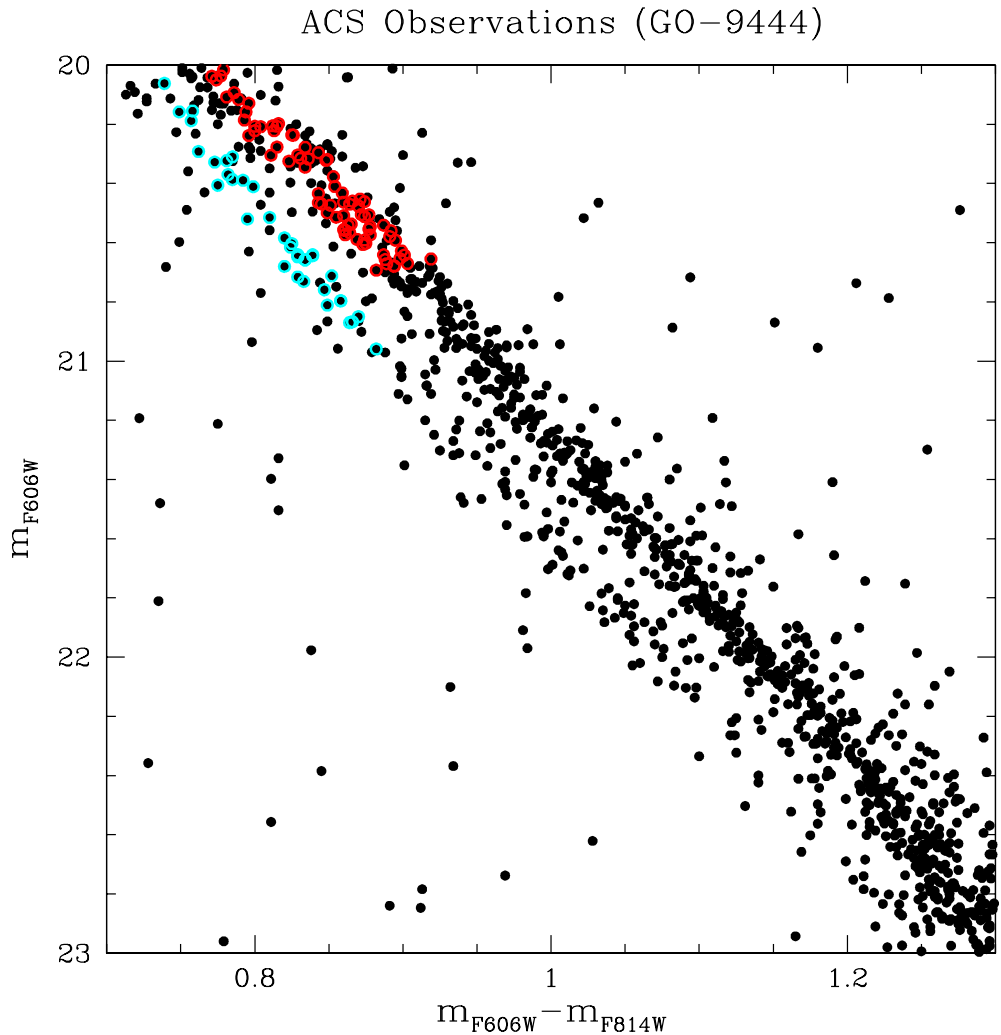}}
   \caption{Color  Magnitude Diagram  of the  $\omega$Cen  Double Main
     Sequence.   In light and  dark grey  circles are  highlighted the
     targets of the FLAMES$+$VLT spectroscopic follow-up.}
   \end{figure*}

Most of the fascinating results on \wcen~ come from the evolved stars,
mainly populating the Red Giant  Branch (RGB). RGB stars have been the
object of a large number of both spectroscopic and photometric surveys
with various groundbased facilities (Norris \& Da Costa, 1995, Pancino
et al.\ 2000, Hilker \&  Richtler 2000) which disclosed the complexity
of the \wcen\ stellar population.

   \begin{figure*}
   \includegraphics[width=12cm]{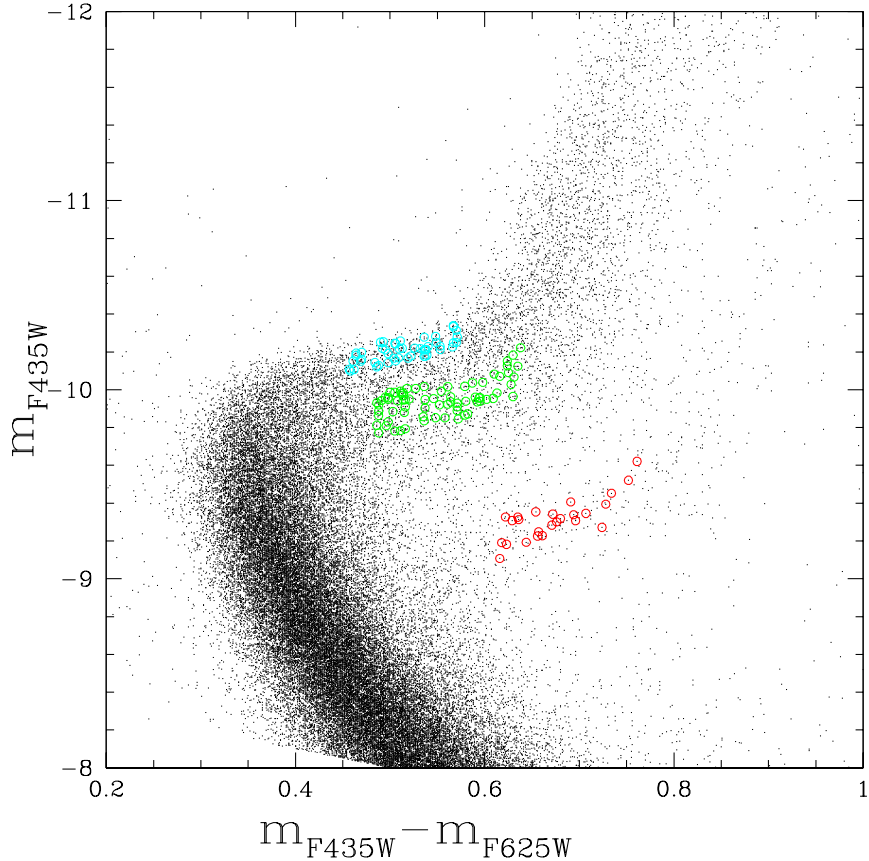}
   \caption{Instrumental Color Magnitude  Diagram around the turn off.
   Circles   highlight  the   target  stars   for   the  spectroscopic
   follow-up. }
   \end{figure*}

More recently, the  high performances of the Wide  Field Channel (WFC)
of the Advanced  Camera for Surveys (ACS) on-board  \hst\ have made it
possible to  explore also the faint unevolved  stellar component, more
efficently  and with  a  much higher  photometric  accuracy than  with
groundbased  facilities. The  new  observations have  made the  \wcen\
stellar  population  scenario even  more  puzzling,  and difficult  to
understand.

\section{The \wcen Double Main Sequence}

Figure 1  shows a  zoom-in of the  color magnitude diagram  (CMD) from
ACS@HST  data of \wcen~  presented by  Bedin et  al.\ (2004a,  see the
paper for a full description of the data base).  The CMD is calibrated
to  the Vega-mag system  (Bedin et  al.\ 2004b,  subm.)  in  the plane
$(m_{\rm  F606W}-m_{\rm  F814W})$  vs.\  $m_{\rm  F606W}$.   This  CMD
confirms the  presence of  a double main  sequence (DMS) in  the color
magnitude  diagram  (CMD)  of  \wcen,  first detected  by  one  of  us
(Anderson 1997, 2002).

The  main  sequence bifurcation  that  we  observe  represents a  real
puzzle, for at least two reasons:

1) The bifurcation itself is  puzzling.  The many detailed photometric
(Pancino et al.\ 2000)  and spectroscopic (Norris, Freeman, \& Mighell
1996) investigations   of  the   RGB  stars   indicate  a   spread  of
metallicities, not two distinct  populations.  The only truly distinct
population seen is the metal-rich component (Pancino et al.\ 2000).

2) The less populous  of our two main sequences (MS)  is the blue one.
This  is even  more difficult  to understand.   Assuming that  all the
stars  in the  two MSs  are members  of \wcen,  any  canonical stellar
models with  canonical chemical abundances  tell us that the  bluer MS
{\it  must}  be  more metal  poor  than  the  red MS.   However,  both
spectroscopic (Norris \& Da  Costa 1995) and photometric (e.g., Hilker
\&  Richtler  2000)  investigations  show  that  the  distribution  in
metallicity  of  the  \wcen~  stars  begins  with  a  peak  at  [Fe/H]
$\!\sim\!-1.6$, and then tails off on the metal-rich side.  

Undoubtely, the  striking result presented raises  more questions than
it  answers.  Various  explanations have  been suggested  in  Bedin et
al.\ 2004, none of them very conclusive.

The  bluest  of the  two  sequences  could  represent an  intermediate
metallicity ($\rm [Fe/H] \!=\!  -1.1$) population of stars formed from
material polluted  by intermediate mass asynptotic  giant branch (AGB)
star ejecta. This material should be helium enhanced. An overabundance
of  helium (Y$>0.3$)  could explain  the blue  color of  the sequence.
Observational  Signatures of this  pollution should  be a  high helium
($\rm  Y\!>\!0.3$) abundance,  high $s$-process  element overabundance
(e.g.\ $\rm [Ba/Fe] \!>\! +1.0$), and possibly high C values.

The  presence of  an  object with  a  metallicity [Fe/H]$\sim-1.0$  in
background,  at $\sim$1.7kpc  from \wcen~  could also  explain  the MS
bifurcation, but this possibility seems rather unlike.

%

\section{Future Work}

Very recently, thanks to a project on ESO Director Discretionary Time,
we acquired new spectroscopic observations at FLAMES@VLT + GIRAFFE and
UVES for 17 blue MS stars and  17 red MS stars.  The targets have been
selected  among  the objects  highlighted  with  light  and dark  grey
circles in Fig.\  1. In addition, several sub  giant branch stars have
been observed, selected among those highlighted in Fig.\ 2.

The purpose of these observations  is to measure the abundance of some
of  the elements  which  may  indicate a  past  pollution episode  (in
particular we  look for Barium overabundance),  providing an essential
ingredient  for  the interpretation  of  the  new  \hst~ results,  and
possibly  for  solving  the  puzzle  still  represented  by  the  star
formation history of \wcen.

Four additional \hst-orbits  during Cycle 13 have been  granted to our
group, to study in detail the  color distribution of the MS stars down
to  the Hydrogen  Burning Limit.  Thanks to  our ability  in measuring
proper motions (Anderson \& King  2000), the new \hst\ data will allow
to investigate any difference in the kinematic properties of these two
MS populations.

These  two new  data  set  will certainly  contribute  in solving  the
stellar population puzzle of \wcen.
  
\begin{acknowledgements}
L.R.B., S.C.,  Y.M., and G.P.\  acknowledge financial support  by MIUR
(PRIN2001,  PRIN2002, and  PRIN2003).  J.A.\  and  I.R.K.\ acknowledge
support by STScI grant GO 9444.
\end{acknowledgements}

\bibliographystyle{aa}

\end{document}